\documentclass[lettersize,journal]{IEEEtran}
\pdfoutput=1
\usepackage{graphicx}
\usepackage{amsmath,amssymb,amsfonts}
\usepackage{algorithmic}

\usepackage{textcomp}
\usepackage{xcolor}
\usepackage[numbers]{natbib}
\usepackage{pifont}

\usepackage{comment}
\usepackage{wrapfig}
\usepackage{colortbl}
\usepackage{listings}
\usepackage[pagebackref,breaklinks,colorlinks,allcolors=cvprblue]{hyperref}
\hypersetup{colorlinks,allcolors=blue}
\usepackage{amssymb}
\usepackage{subcaption}
\usepackage{tikz}
\usetikzlibrary{math}
\usepackage{threeparttable}
\usepackage{booktabs}
\usepackage{multirow}
\usepackage{graphicx} 
\usepackage{float} 
\usepackage[utf8]{inputenc} 
\usepackage[T1]{fontenc}    
\usepackage{hyperref}       
\usepackage{url}            
\usepackage{booktabs}       
\usepackage{amsfonts}       
\usepackage{nicefrac}       
\usepackage{microtype}      
\usepackage{xcolor}         
\usepackage{rotating}
\usepackage{tcolorbox}
\usepackage{makecell}

\newcommand{\answerTODO}[1][]{\textcolor{red}{\bf [TODO]}}

\definecolor{dkgreen}{rgb}{0,0.6,0}
\definecolor{gray}{rgb}{0.5,0.5,0.5}
\definecolor{mauve}{rgb}{0.58,0,0.82}
\definecolor{almond}{rgb}{0.94, 0.87, 0.8}
\definecolor{babyblueeyes}{rgb}{0.63, 0.79, 0.95}
\definecolor{beige}{rgb}{0.96, 0.96, 0.86}
\definecolor{anti-flashwhite}{rgb}{0.95, 0.95, 0.96}
\definecolor{darkblue}{rgb}{0.03, 0.27, 0.49}
\definecolor{darkgreen}{rgb}{0.01, 0.75, 0.24}
\definecolor{darkred}{rgb}{0.76, 0.23, 0.13}
\definecolor{light-gray}{gray}{0.92}
\graphicspath{{fig/}}

\title{A Plug-and-Play Temporal Normalization Module for \\ Robust Remote Photoplethysmography}

\author{
\IEEEauthorblockN{Kegang Wang,
Jiankai Tang, Yantao Wei, Mingxuan Liu, Xin Liu, Yuntao Wang
}\\
}

\begin{document}
\maketitle
\begin{abstract}
Remote photoplethysmography (rPPG) extracts PPG signals from subtle color changes in facial videos, showing strong potential for health applications. However, most rPPG methods rely on intensity differences between consecutive frames, missing long-term signal variations affected by motion or lighting artifacts, which reduces accuracy. This paper introduces Temporal Normalization (TN), a flexible plug-and-play module compatible with any end-to-end rPPG network architecture. By capturing long-term temporally normalized features following detrending, TN effectively mitigates motion and lighting artifacts, significantly boosting the rPPG prediction performance. When integrated into four state-of-the-art rPPG methods, TN delivered performance improvements ranging from 34.3\% to 94.2\% in heart rate measurement tasks across four widely-used datasets. Notably, TN showed even greater performance gains in smaller models. We further discuss and provide insights into the mechanisms behind TN’s effectiveness.

\end{abstract}

\section{Introduction}

Remote photoplethysmography (rPPG) is a non-contact method that extracts physiological signals using the principles of the Shafer Reflection Model (SRM) \cite{shafer, wang2016algorithmicprinciple} and the variable absorption spectra of hemoglobin based on oxygen levels \cite{edwards1993measurement}. The diffusely reflected light from beneath the skin, carrying periodic cardiac information, is altered by the oxygen content in the blood. Although these signals are subtle and imperceptible to the human eye, they can be captured by cameras \cite{wu2012eulerianmotion}. rPPG technology enables the extraction of blood volume pulse (BVP) signals and, consequently, critical cardiac indices such as heart rate (HR) \cite{liu2022rppgtoolbox, mcduff2019iphys}, heart rate variability (HRV) \cite{wang2023physbench}, and blood oxygen saturation (SpO2) \cite{liu2024summit, spo2S2024}. Influenced by physical activity, emotional states, and health conditions, rPPG has potential for applications in healthcare, fitness, and affective computing \cite{tang2024camera, tang2023alpha}.

The initial rPPG methods used unsupervised signal processing algorithms based on prior assumptions about skin color, ambient light, motion, and periodic prior knowledge of cardiac signals. Techniques such as blind source separation \cite{ICA, ICA1, pca} and spatio-temporal skin color information extraction algorithms \cite{verkruysse2008green, hulsbusch2008greenred, chrom, pbv2014, pos, samc} lacked robustness in noisy real-world scenarios due to their reliance on these assumptions.

\begin{figure}
    \includegraphics[width=0.5\textwidth]{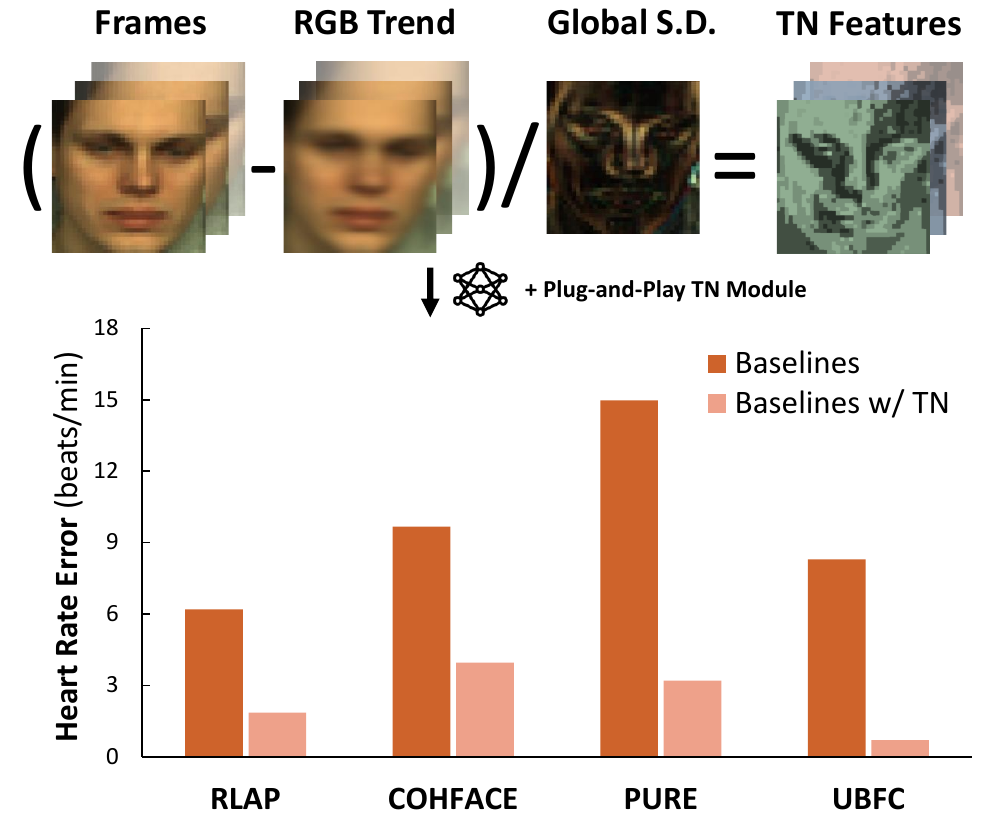}
    \caption{\textbf{TN module improves rPPG performance.} The Plug-and-Play TN module detrend pixels along the temporal dimension and divide by the Standard Deviation (S.D.), demonstrates a strong capacity for noise suppression, leading to an average error reduction of 75\% for four baselines (TS-CAN\cite{mttscan}, PhysNet\cite{physnet}, EffPhys-C\cite{efficientphys}, and PhysFormer\cite{physformer}), trained on the MMPD\cite{tang2023mmpd} dataset.} 
    \label{fig:teaser}
\end{figure}

In contrast, the first neural rPPG model, DeepPhys \cite{deepphys}, integrated differential features from the outset to enhance signal detection and reduce noise. This approach, using differential-normalized features as network inputs, has been widely adopted in models such as TS-CAN \cite{mttscan}, EfficientPhys \cite{efficientphys}, BigSmall \cite{narayanswamy2023bigsmall}, and PhysMamba \cite{physmambayu}. However, the pre-processing required to acquire these features introduces significant computational overhead.

\begin{figure*}[t!]
    \centering
    \includegraphics[width=\textwidth]{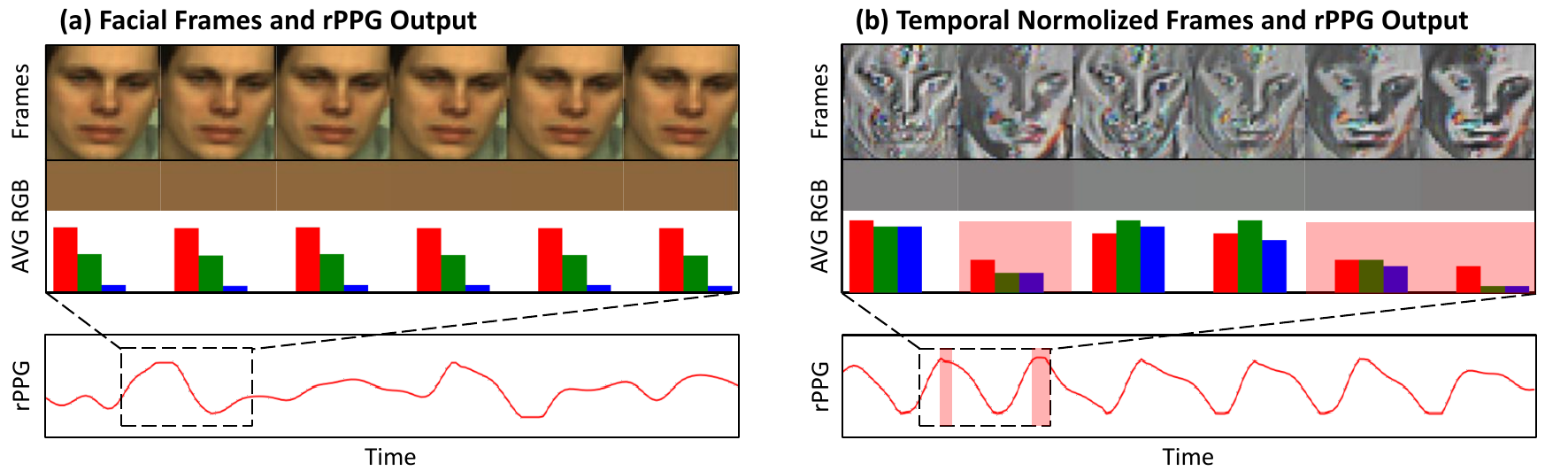}
    \vspace{-0.2cm}
    \caption{\textbf{Visualization of temporal normalization (TN) module.} (a)Without TN, the difference between facial frames and the average RGB values is slight, resulting in noisy prediction. (b)After adding the TN module, the temporal features and the difference of averages are amplified, leading to strong periodic patterns corresponding to the peaks of the BVP waveform.}
    \label{fig:tnvis}
\end{figure*}

To streamline this process, end-to-end models have been developed that include internal differential modules. AutoHR pioneered the use of Temporal Difference Convolution (TDC) \cite{autohrtdc}; its principle involves calculating the average of preceding and succeeding frames along with the difference with the intermediate frame. Nevertheless, manually processing these features incurs substantial computational overhead. Thus, a novel convolutional kernel was proposed, equivalent to manual differential operations, but accomplishable through convolution operators, thus enhancing computational efficiency while preserving the model’s end-to-end input-output. The TDC module has been adopted in numerous subsequent works\cite{physformer, physformer++,  Zou2024RhythmFormerER, physmambayu, 3ddiffattn}, with 3DLDC \cite{huang2024ddrppgnetdeinterferingdescriptivefeature} serving as a generalized version of TDC. In works like PhysMamba \cite{Yan2024PhysMambaSS} and RhythmMamba \cite{Zou2024RhythmMambaFR}, the features traverse a stem layer equipped with a differential module, dividing the video input into two branches: an appearance branch and a differential branch, which are subsequently merged before being processed in Mamba blocks. Differential Temporal Convolution (DTC) \cite{talos} employs multi-level derivatives to extract differential features, focusing on one-dimensional temporal features and thereby reducing computational complexity. Differential features can also be extracted through spatiotemporal convolutions\cite{3Ddil, rtrppg, sun2022contrast, sun2024contrastp}, where the temporal dimension of the convolution kernel enables capturing small-scale temporal features. This capability allows the detection of subtle differences between frames, which is the operational principle of some end-to-end models based on convolutional algorithms. Additionally, approaches using Spatiotemporal Maps (STMap) \cite{synrhythm, rhythmnet, cvd, dualgan, transrppg, rsrppg} craft three-dimensional spatiotemporal features into two-dimensional feature maps processed through convolutional networks. This method reduces the dimensionality of network input but increases preprocessing complexity.

Despite advancements, differential features present inherent challenges.  They emphasize minor local chromatic changes but lack a noise-filtering mechanism, leading to the amplification of even minimal noise signals. Given that heart rate (HR) frequencies typically hover around 60 BPM, describing one heartbeat cycle generally requires about 30 frames, and differential features fail to encompass a complete heartbeat cycle. This poses two challenges: \textbf{how to filter noise from differential features and how to perform long-term association modeling to capture comprehensive cardiac cyclical features.}

This paper revisits the foundation of neural rPPG models: the Shafer Reflectance Model (SRM), to propose a novel feature extraction method: Temporal Normalization (TN). TN is a plug-and-play module capable of amplifying subtle chromatic features. In addition to enhancing rPPG features, it offers significant advantages over differential methods, such as robust environmental noise suppression and the ability to extract global features for long-term correlation modeling. These capabilities enable TN to address current major challenges effectively in a plug-and-play manner.

Our key contributions are as follows:

\begin{itemize}

\item We introduce a plug-and-play module, Temporal Normalization (TN), that seamlessly integrates into any end-to-end rPPG models, without requiring additional parameters. It significantly enhanced the model's robustness, resulting in an up to 94.2\% reduction in the Mean Absolute Error (MAE) for cross-dataset experiments on four baselines.



\item The TN module facilitates the training of models on highly noisy datasets, addressing the previously encountered issue of model non-convergence. Models trained on MMPD\cite{tang2023mmpd} achieved a clinical-level (MAE below 2) performance on the other four datasets for the first time.

\item The TN module can endow models with low-cost long-term associations, allowing for the extraction of spatiotemporal rPPG features with minimal parameters. Research on PhysNet\cite{physnet} variants demonstrates that after reducing the parameter count by 98.8\%, PhysNet-TN still performs better than the original version.


\end{itemize}
\section{Princple and Method}
\label{sec: real optical model}
\begin{figure*}
    \centering
    \includegraphics[width=\textwidth]{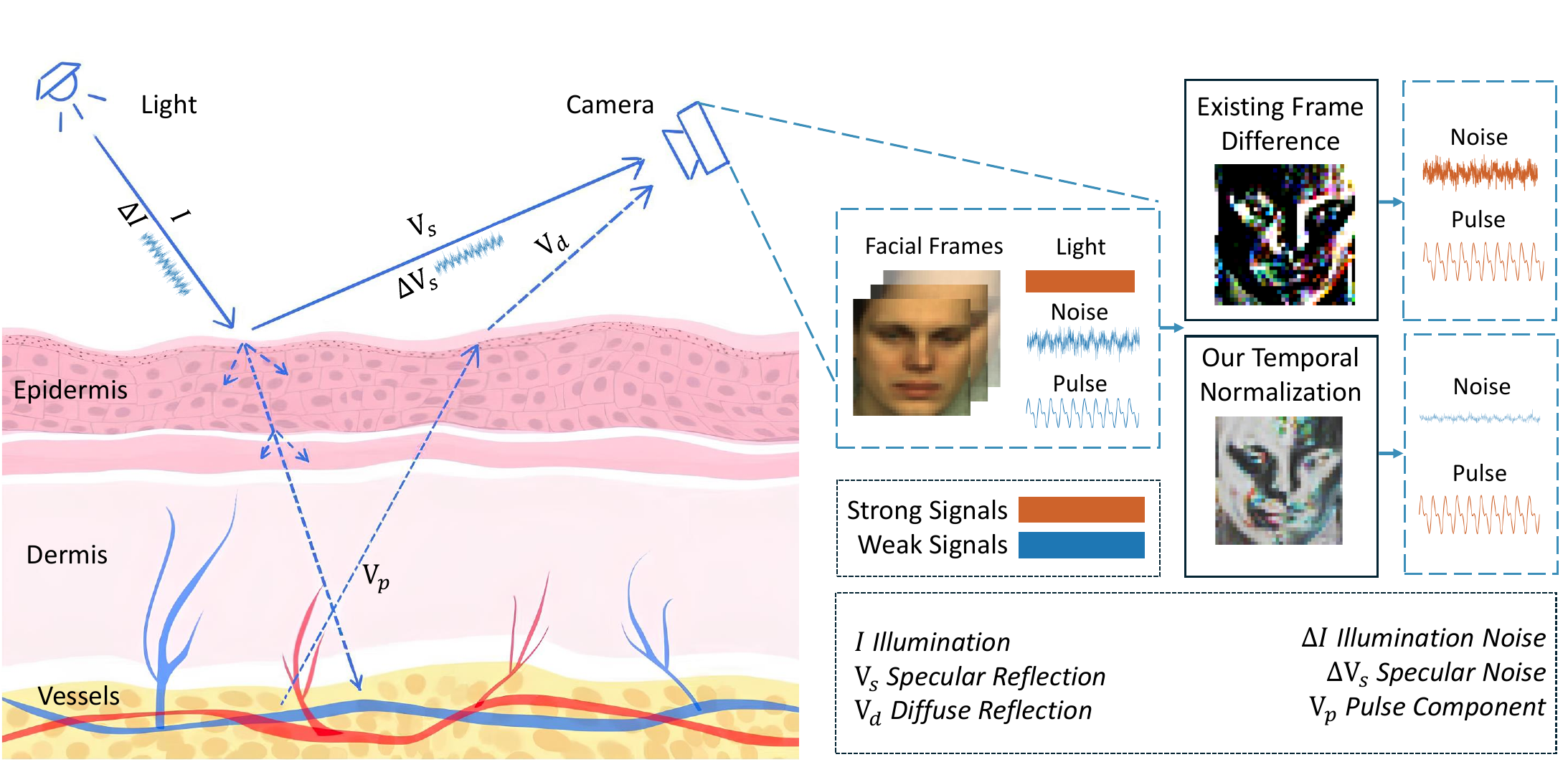}
    \caption{\textbf{TN enhances the pulse signal component and reduces optical noise. }According to a real-world optical model, the camera primarily captures stable light along with periodic noise and pulse. Traditional difference methods eliminate the light component but also amplify noise and the pulse. Our TN module effectively weakens the light and noise while enhancing the desired pulse signal. }
    \label{fig:priciple}
\end{figure*}
Based on the Shafer reflectance model, we have transferred differential features to normalized features to achieve greater robustness. Section \ref{sec: real optical model} introduces the fundamental optical model for the rPPG task, while Section \ref{sec: differential method} discusses adjacent differential frame features and their limitations. Section \ref{sec: temporal method} presents our proposed temporal normalization module and its invariance to noise.


\subsection{Optical Model}

The motivation for proposing the TN module stems from a re-evaluation of the Shafer Reflectance Model (SRM)\cite{shafer}. The schematic diagram of the SRM optical model of rPPG is illustrated in Figure \ref{fig:priciple}. The signal captured by the camera is divided into two components: the specular reflection component of the surface of the skin and the diffuse reflection component from beneath the skin, denoted \(\boldsymbol{v_s}\) and \(\boldsymbol{v_d}\), respectively. The diffuse reflection component \(\boldsymbol{v_d}\) is affected by changes in blood volume, with the blood volume factor represented as \(\boldsymbol{v_p}\).

Therefore, the ideal optical model of the rPPG signal can be expressed by Equation \eqref{optical1}:

\begin{equation}
    \begin{aligned}
    \boldsymbol{C}_I(v) &= \boldsymbol{I} \cdot (\boldsymbol{v_s}+\boldsymbol{v_d} \cdot (1+\boldsymbol{v_p}(v)))
    \label{optical1}
    \end{aligned}
\end{equation}

In real-world scenarios, remote photoplethysmography (rPPG) tasks are influenced by various interfering factors. These noise components can be hypothetically divided into two parts, such as the motion noise present in the specular reflection component \(_{\Delta}\boldsymbol{v_s}\) and the illumination variation present in the light source \(_{\Delta}\boldsymbol{I}\). 
These components can be expressed in Equation \eqref{nonidea} as follows:

\begin{equation}
    \begin{aligned}
    \boldsymbol{C}_R(v) &= (\boldsymbol{I}+_{\Delta}\boldsymbol{I}) \cdot (\boldsymbol{v_s}+_{\Delta}\boldsymbol{v_s}+\boldsymbol{v_d} \cdot (1+\boldsymbol{v_p}(v)))
    \label{nonidea}
    \end{aligned}
\end{equation}

Calculating the impact of noise components on the optical signal, Equation \eqref{optical1} is subtracted from Equation \eqref{nonidea}, resulting in:

\begin{equation}
\begin{aligned}
    _{\Delta}\boldsymbol{C}(v) 
    &= \boldsymbol{I} \cdot _{\Delta}\boldsymbol{v_s} + _{\Delta}\boldsymbol{I} \cdot (\boldsymbol{v_s}+_{\Delta}\boldsymbol{v_s} +\boldsymbol{v_d} \cdot (1+\boldsymbol{v_p}(v))
    \label{cdelta}
\end{aligned}
\end{equation}

Assuming that the intensity of the noise components is similar to the intensity of the rPPG signal and is significantly less than the intensity of illumination, it can be represented by the following equation:

\begin{equation}
    \boldsymbol{I} \sim \boldsymbol{v_s} \sim \boldsymbol{v_d} \gg _{\Delta}\boldsymbol{I} \sim _{\Delta}\boldsymbol{v_s} \sim \boldsymbol{v_p}(v)
    \label{order}
\end{equation} 

Thus, Equation \eqref{cdelta} can be approximated as:

\begin{equation}
\begin{aligned}
    _{\Delta}\boldsymbol{C}(v) &\approx \boldsymbol{I} \cdot _{\Delta}\boldsymbol{v_s} + _{\Delta}\boldsymbol{I} \cdot (\boldsymbol{v_s}+\boldsymbol{v_d})\\&
    \sim \boldsymbol{v_p}(v)
    \label{nonideaappr}
\end{aligned}
\end{equation}

At this point, the impact of the noise term \(_{\Delta}\boldsymbol{C}(v)\) on the rPPG signal is similar to that of the blood volume signal. It describes the core challenge in current rPPG tasks: how to filter out the noise components from the rPPG signal, retaining only the blood volume component.

\subsection{Existing Differential Neural Models} 
\label{sec: differential method}
The general neural network model's learning target is to reconstruct the BVP signal from subtle color changes between two consecutive video frames. Such a model can be termed a "differential model." If the input and output of the model are modified to be the first derivative of blood volume, Equation \eqref{nonidea} can be rewritten as:

\begin{equation}
        \boldsymbol{C^{'}}(v) = (\boldsymbol{I}+_{\Delta}\boldsymbol{I}) \cdot \boldsymbol{v_d} \cdot \boldsymbol{{v_p}^{'}}(v)
        \label{diff1}
\end{equation}

Similar to Equation \eqref{cdelta}, calculate the impact of noise on the differential signal:

\begin{equation}
\begin{aligned}
_{\Delta}\boldsymbol{C^{'}}(v) &= _{\Delta}\boldsymbol{I} \cdot \boldsymbol{v_d} \cdot \boldsymbol{{v_p}^{'}}(v)\\
&< _{\Delta}\boldsymbol{C}(v)
\label{cddelta}
\end{aligned}
\end{equation}

Equation \eqref{cddelta} describes the approach of general models, which involves filtering noise terms through the differential signal. At this point, the magnitude of the noise component \(_{\Delta}\boldsymbol{C^{'}}(v)\) is smaller than that of the Equation \eqref{nonideaappr}, allowing the model to learn the BVP signal. However, it is evident that the model's output will still be influenced by changes in illumination. 

\subsection{Temporal Normalized Neural Models} 
\label{sec: temporal method}

To overcome the limitations of the previous method to eliminate incomplete noise, we propose a plug-and-play module for the remote photoplethysmography (rPPG) task, termed Temporal Normalization (TN). TN uses temporally normalized features to replace the previous differential features, which comprises two steps in the module design: (1) detrending, and (2) normalization of the temporal dimension. 

Detrending is performed using linear regression based on the least squares method, as demonstrated in Equation \eqref{detrend}.

\begin{equation}
    \Tilde{P}_{i, j}(t) = P_{i, j}(t) - (\hat{\alpha}_{i, j}t+\hat{\beta}_{i, j})
    \label{detrend}
\end{equation}

Where \(\hat{\alpha}_{i, j}\) represents the slope of the \( (i, j) \)th pixel, \(\hat{\beta}_{i, j}\) represents the intercept of the \( (i, j) \)th pixel, and \(\Tilde{P}_{i, j}(t)\) represents the detrended value of the \( (i, j) \)th pixel. 
After detrending, apply RMS-Norm on the time axis, as illustrated in Equation \eqref{normalization}.

\begin{equation}
    \hat{P}_{i, j}(t) = \frac{\Tilde{P}_{i, j}(t)}{\sqrt{\frac{1}{T}\sum\limits_{t=0}^{T}\Tilde{P}^2_{i, j, t}+\epsilon}}
    \label{normalization}
\end{equation} 

Where \(\epsilon\) is a small constant. 

Consider normalizing the rPPG signal and reformulating Equation \eqref{optical1}, the normalized signal \(\boldsymbol{\hat{C}}(v)\) can be represented as:

\begin{equation}
    \begin{aligned}
    \boldsymbol{\hat{C}}(v) &= \frac{\boldsymbol{C}(v)-\boldsymbol{\Bar{C}}}{\sigma(\boldsymbol{C})}\\
    &= \frac{\boldsymbol{I}\cdot\boldsymbol{v_d} \cdot (\boldsymbol{v_p}(v)-\boldsymbol{\Bar{v_p}})}{\sigma(\boldsymbol{I}\cdot\boldsymbol{v_d}\cdot\boldsymbol{v_p})}
    \label{cnorm}
    \end{aligned}
\end{equation} 

Where \(\boldsymbol{\Bar{v_p}}\) is the mean value of blood volume signal, which is a periodic pulse signal, according to Equation \eqref{detrend}, its mean is zero after detrending, the equation can be simplified to: 

\begin{equation} 
    \begin{aligned}
    \boldsymbol{\hat{C}}(v) &= \frac{\boldsymbol{v_p}(v)}{\sigma(\boldsymbol{v_p})}
    \label{cnorm1}
    \end{aligned}
\end{equation}

Similar to Equation \eqref{cdelta}, calculate the impact of noise on the normalized signal:

\begin{equation}
\begin{aligned} 
    _{\Delta}\boldsymbol{\hat{C}}(v) &= 0 
    \label{cnorm2} 
\end{aligned} 
\end{equation} 

\begin{figure}[ht]
    \centering
    \includegraphics[width=82.5mm]{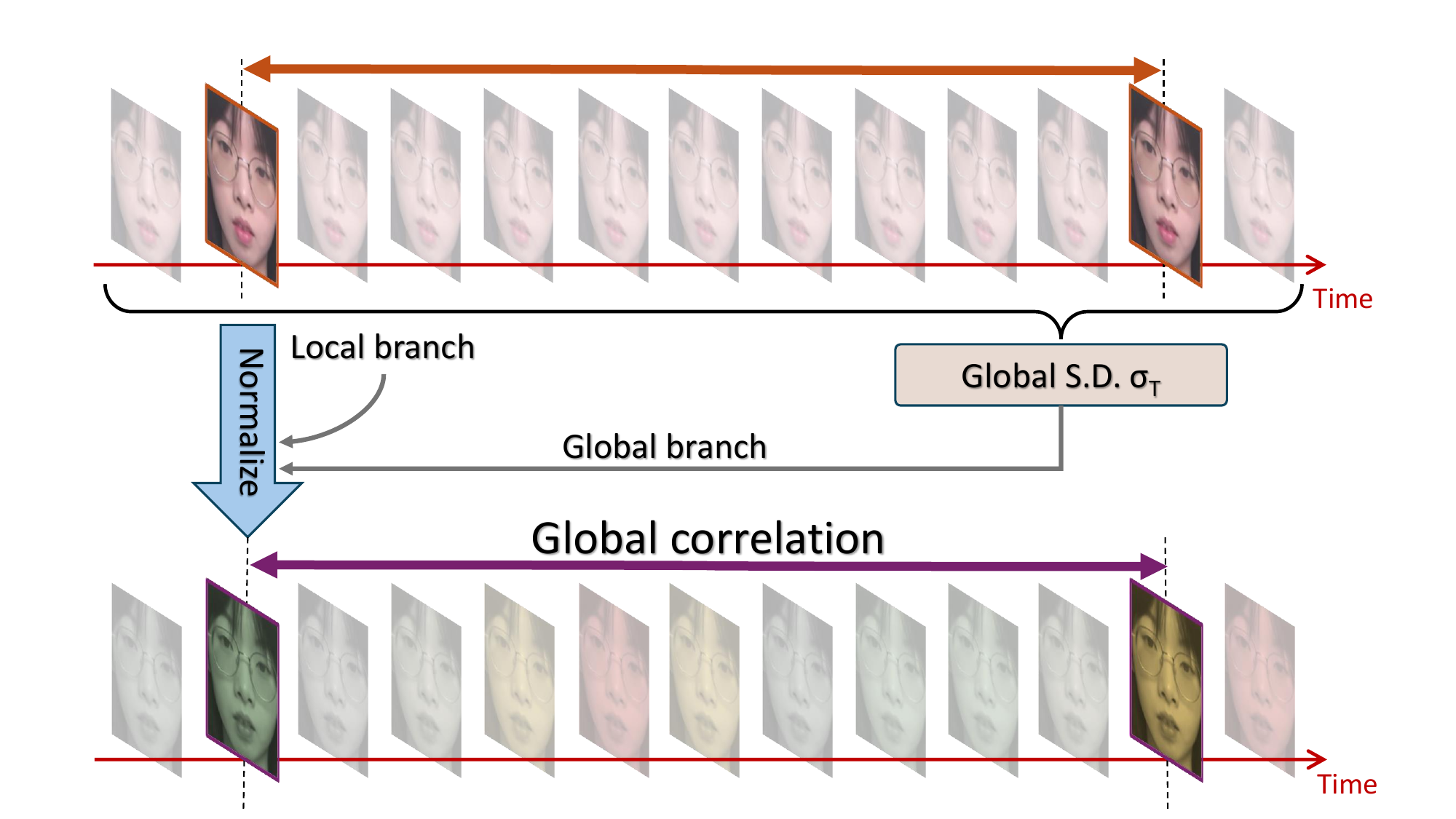}
    \caption{\textbf{TN establishes the global correlation across long frames.} The temporal normalization features are integrated through two branches: the local frame features and the global standard deviation features. These jointly compute the normalized features, enhancing the global correlation of the features.}
    \label{fig:global}
\end{figure}

\section{Experiments}  

\begin{table*}[t!]
\caption{Cross-testing on the MMPD\cite{tang2023mmpd}, COHFACE\cite{cohface}, PURE\cite{pure}, and UBFC\cite{ubfcrppg} (trained on RLAP\cite{wang2023physbench}). \textbf{Bold}: The best result.}
\label{resultrlap}
\centering
\setlength{\tabcolsep}{0pt} 
\scalebox{1.1}{
\begin{threeparttable}
\begin{tabular}{lccccccccccccc}
  \toprule
  \multirow{3}{*}{\bf{Method}} & \multirow{3}{*}{\bf{w/TN}} & \multicolumn{3}{c}{\bf{MMPD\cite{tang2023mmpd}}} & \multicolumn{3}{c}{\bf{COHFACE\cite{cohface}}} & \multicolumn{3}{c}{\bf{PURE\cite{pure}}} & \multicolumn{3}{c}{\bf{UBFC\cite{ubfcrppg}}} \\
  \cmidrule(lr){3-5}\cmidrule(lr){6-8}\cmidrule(lr){9-11}\cmidrule(lr){12-14}
  & & MAE↓ & RMSE↓ & $\rho$↑ & MAE↓ & RMSE↓ & $\rho$↑ & MAE↓ & RMSE↓ & $\rho$↑ & MAE↓ & RMSE↓ & $\rho$↑ \\
  \midrule
   \multirow{3}{*}{TN-Net (Ours)} & \ding{55} & 31.70 & 35.90 & 0.035 & 21.41 & 25.31 & -0.25 & 17.83 & 28.21 & 0.049 & 38.04 & 42.85 & 0.063 \\
     &  \cellcolor{light-gray} \checkmark & \cellcolor{light-gray}\bf{6.620} & \cellcolor{light-gray}\bf{10.46} & \cellcolor{light-gray}\bf{0.734} & \cellcolor{light-gray}2.514 \cellcolor{light-gray}& 5.672 \cellcolor{light-gray}& 0.875 \cellcolor{light-gray}& \bf{0.232} \cellcolor{light-gray}& 0.557 \cellcolor{light-gray}& \bf{1.000} \cellcolor{light-gray}& 0.448 \cellcolor{light-gray}& 0.701 \cellcolor{light-gray}& \cellcolor{light-gray}\bf{0.999} \\
          & \textsc{Delta} & \textcolor{darkgreen}{\small \texttt{-}\textbf{79.1\%}} & \textcolor{darkgreen}{\small \texttt{-}\textbf{70.9\%}} & \textcolor{darkgreen}{\small \texttt{+}\textbf{1,997\%}} & \textcolor{darkgreen}{\small \texttt{-}\textbf{88.3\%}} & \textcolor{darkgreen}{\small \texttt{-}\textbf{77.6\%}} & \textcolor{darkgreen}{\small \texttt{+}\textbf{460\%}} & \textcolor{darkgreen}{\small \texttt{-}\textbf{98.7\%}} & \textcolor{darkgreen}{\small \texttt{-}\textbf{98.0\%}} & \textcolor{darkgreen}{\small \texttt{+}\textbf{1,941\%}} & \textcolor{darkgreen}{\small \texttt{-}\textbf{98.8\%}} & \textcolor{darkgreen}{\small \texttt{-}\textbf{98.4\%}} & \textcolor{darkgreen}{\small \texttt{+}\textbf{1,485\%}} \\
  \midrule
  \multirow{3}{*}{TS-CAN\cite{mttscan}}   & \ding{55} & 10.19 & 15.16 & 0.471 & 6.580 & 9.714 & 0.610 & 3.902 & 7.045 & 0.959 & 0.904 & 1.658 & 0.996 \\
           & \cellcolor{light-gray} \checkmark & \cellcolor{light-gray}9.127 & \cellcolor{light-gray}14.23 & \cellcolor{light-gray}0.583 & \cellcolor{light-gray}2.536 & \cellcolor{light-gray}6.095 & \cellcolor{light-gray}0.860 & \cellcolor{light-gray}0.281 & \cellcolor{light-gray}0.728 & \cellcolor{light-gray}\bf{1.000} & \cellcolor{light-gray}0.561 & \cellcolor{light-gray}0.806 & \cellcolor{light-gray}\bf{0.999} \\
           & \textsc{Delta} & \textcolor{darkgreen}{\small \texttt{-}\textbf{10.4\%}} & \textcolor{darkgreen}{\small \texttt{-}\textbf{6.1\%}} & \textcolor{darkgreen}{\small \texttt{+}\textbf{23.7\%}} & \textcolor{darkgreen}{\small \texttt{-}\textbf{61.5\%}} & \textcolor{darkgreen}{\small \texttt{-}\textbf{37.2\%}} & \textcolor{darkgreen}{\small \texttt{+}\textbf{40.8\%}} & \textcolor{darkgreen}{\small \texttt{-}\textbf{92.8\%}} & \textcolor{darkgreen}{\small \texttt{-}\textbf{89.7\%}} & \textcolor{darkgreen}{\small \texttt{+}\textbf{4.3\%}} & \textcolor{darkgreen}{\small \texttt{-}\textbf{37.9\%}} & \textcolor{darkgreen}{\small \texttt{-}\textbf{51.4\%}} & \textcolor{darkgreen}{\small \texttt{+}\textbf{0.3\%}} \\
  \midrule
  \multirow{3}{*}{PhysNet\cite{physnet}}  & \ding{55} & 11.17 & 16.41 & 0.418 & 7.380 & 9.510 & 0.559 & 5.145 & 13.06 & 0.839 & 0.655 & 1.034 & 0.998 \\
           &\cellcolor{light-gray} \checkmark &\cellcolor{light-gray} 9.386 &\cellcolor{light-gray} 14.81 &\cellcolor{light-gray} 0.566 &\cellcolor{light-gray} \bf{2.240} &\cellcolor{light-gray} \bf{5.262} &\cellcolor{light-gray} \bf{0.893} &\cellcolor{light-gray} 0.363 &\cellcolor{light-gray} 1.165 &\cellcolor{light-gray} 0.999 &\cellcolor{light-gray} 0.535 &\cellcolor{light-gray} 0.784 &\cellcolor{light-gray} \bf{0.999} \\
           & \textsc{Delta} & \textcolor{darkgreen}{\small \texttt{-}\textbf{16.0\%}} & \textcolor{darkgreen}{\small \texttt{-}\textbf{9.8\%}} & \textcolor{darkgreen}{\small \texttt{+}\textbf{35.4\%}} & \textcolor{darkgreen}{\small \texttt{-}\textbf{69.7\%}} & \textcolor{darkgreen}{\small \texttt{-}\textbf{44.7\%}} & \textcolor{darkgreen}{\small \texttt{+}\textbf{59.8\%}} & \textcolor{darkgreen}{\small \texttt{-}\textbf{92.9\%}} & \textcolor{darkgreen}{\small \texttt{-}\textbf{91.1\%}} & \textcolor{darkgreen}{\small \texttt{+}\textbf{19.1\%}} & \textcolor{darkgreen}{\small \texttt{-}\textbf{18.3\%}} & \textcolor{darkgreen}{\small \texttt{-}\textbf{24.2\%}} & \textcolor{darkgreen}{\small \texttt{+}\textbf{0.1\%}} \\
  \midrule
  \multirow{3}{*}{EFFPhys-C\cite{efficientphys}} & \ding{55} & 9.197 & 13.73 & 0.555 & 4.243 & 6.962 & 0.792 & 2.483 & 5.481 & 0.975 & 0.812 & 1.707 & 0.995 \\
           &\cellcolor{light-gray} \checkmark &\cellcolor{light-gray} 10.67 &\cellcolor{light-gray} 16.48 &\cellcolor{light-gray} 0.459 &\cellcolor{light-gray} 2.764 &\cellcolor{light-gray} 5.940 &\cellcolor{light-gray} 0.870 &\cellcolor{light-gray} 0.248 &\cellcolor{light-gray} \bf{0.409} &\cellcolor{light-gray} \bf{1.000} &\cellcolor{light-gray} 0.571 &\cellcolor{light-gray} 0.893 &\cellcolor{light-gray} \bf{0.999} \\
           & \textsc{Delta} & \textcolor{darkgray}{\small \texttt{+}\textbf{16.0\%}} & \textcolor{darkgray}{\small \texttt{+}\textbf{20.0\%}} & \textcolor{darkgray}{\small \texttt{-}\textbf{17.3\%}} & \textcolor{darkgreen}{\small \texttt{-}\textbf{34.9\%}} & \textcolor{darkgreen}{\small \texttt{-}\textbf{14.7\%}} & \textcolor{darkgreen}{\small \texttt{+}\textbf{9.8\%}} & \textcolor{darkgreen}{\small \texttt{-}\textbf{90.0\%}} & \textcolor{darkgreen}{\small \texttt{-}\textbf{92.5\%}} & \textcolor{darkgreen}{\small \texttt{+}\textbf{2.6\%}} & \textcolor{darkgreen}{\small \texttt{-}\textbf{29.7\%}} & \textcolor{darkgreen}{\small \texttt{-}\textbf{47.7\%}} & \textcolor{darkgreen}{\small \texttt{+}\textbf{0.4\%}} \\
  \midrule
  \multirow{3}{*}{PhysFormer\cite{physformer}} & \ding{55} & 12.82 & 18.85 & 0.282 & 7.767 & 11.84 & 0.424 & 6.060 & 12.38 & 0.853 & 0.503 & 0.745 & \bf{0.999} \\
            &\cellcolor{light-gray} \checkmark &\cellcolor{light-gray} 8.136 &\cellcolor{light-gray} 13.22 &\cellcolor{light-gray} 0.586 &\cellcolor{light-gray} 2.597 &\cellcolor{light-gray} 6.010 &\cellcolor{light-gray} 0.856 &\cellcolor{light-gray} 0.238 &\cellcolor{light-gray} 0.590 &\cellcolor{light-gray} \bf{1.000} &\cellcolor{light-gray} \bf{0.446} &\cellcolor{light-gray} \bf{0.673} &\cellcolor{light-gray} \bf{0.999} \\
            & \textsc{Delta} & \textcolor{darkgreen}{\small \texttt{-}\textbf{36.5\%}} & \textcolor{darkgreen}{\small \texttt{-}\textbf{29.9\%}} & \textcolor{darkgreen}{\small \texttt{+}\textbf{107.8\%}} & \textcolor{darkgreen}{\small \texttt{-}\textbf{66.6\%}} & \textcolor{darkgreen}{\small \texttt{-}\textbf{49.2\%}} & \textcolor{darkgreen}{\small \texttt{+}\textbf{102.0\%}} & \textcolor{darkgreen}{\small \texttt{-}\textbf{96.1\%}} & \textcolor{darkgreen}{\small \texttt{-}\textbf{95.2\%}} & \textcolor{darkgreen}{\small \texttt{+}\textbf{17.2\%}} & \textcolor{darkgreen}{\small \texttt{-}\textbf{11.3\%}} & \textcolor{darkgreen}{\small \texttt{-}\textbf{9.7\%}} & \textcolor{darkgreen}{\small \texttt{-}\textbf{0.0\%}} \\
  \bottomrule
\end{tabular}
     \begin{tablenotes}[flushleft]
     \item \textbf{MAE}: Mean Absolute Error, \textbf{RMSE}: Root Mean Square Error, \textbf{$\rho$}: Pearson Correlation Coefficient.
     \end{tablenotes}
\end{threeparttable}
}
\end{table*}

\begin{table*}[ht]
\caption{Cross-testing on the RLAP\cite{wang2023physbench}, COHFACE\cite{cohface}, PURE\cite{pure}, and UBFC\cite{ubfcrppg} (trained on MMPD\cite{tang2023mmpd}). \textbf{Bold}: The best result.}
\label{resultmmpd}
\centering
\setlength{\tabcolsep}{0pt} 
\scalebox{1.1}{
\begin{threeparttable}
\begin{tabular}{lccccccccccccc}
  \toprule
  \multirow{3}{*}{\bf{Method}} & \multirow{3}{*}{\bf{w/TN}} & \multicolumn{3}{c}{\bf{RLAP\cite{wang2023physbench}}} & \multicolumn{3}{c}{\bf{COHFACE\cite{cohface}}} & \multicolumn{3}{c}{\bf{PURE\cite{pure}}} & \multicolumn{3}{c}{\bf{UBFC\cite{ubfcrppg}}} \\
  \cmidrule(lr){3-5}\cmidrule(lr){6-8}\cmidrule(lr){9-11}\cmidrule(lr){12-14}
  & & MAE↓ & RMSE↓ & $\rho$↑ & MAE↓ & RMSE↓ & $\rho$↑ & MAE↓ & RMSE↓ & $\rho$↑ & MAE↓ & RMSE↓ & $\rho$↑ \\
  \midrule
  \multirow{3}{*}{TN-Net (Ours)} & \ding{55} & 25.93 & 27.99 & 0.056 & 19.93 & 23.35 & 0.006 & 19.51 & 28.55 & 0.062 & 44.08 & 47.55 & 0.086 \\
            &\cellcolor{light-gray}  \checkmark &\cellcolor{light-gray} \bf{1.114} &\cellcolor{light-gray} \bf{2.337} &\cellcolor{light-gray} \bf{0.975} &\cellcolor{light-gray} 3.037 &\cellcolor{light-gray} 7.018 &\cellcolor{light-gray} 0.809 &\cellcolor{light-gray} \bf{0.622} &\cellcolor{light-gray} \bf{2.296} &\cellcolor{light-gray} \bf{0.995} &\cellcolor{light-gray} 0.546 &\cellcolor{light-gray} 0.815 &\cellcolor{light-gray} \bf{0.999} \\
            & \textsc{Delta} & \textcolor{darkgreen}{\small \texttt{-}\textbf{95.7\%}} & \textcolor{darkgreen}{\small \texttt{-}\textbf{91.7\%}} & \textcolor{darkgreen}{\small \texttt{+}\textbf{1,641\%}} & \textcolor{darkgreen}{\small \texttt{-}\textbf{84.8\%}} & \textcolor{darkgreen}{\small \texttt{-}\textbf{69.9\%}} & \textcolor{darkgreen}{\small \texttt{+}\textbf{13,383\%}} & \textcolor{darkgreen}{\small \texttt{-}\textbf{96.8\%}} & \textcolor{darkgreen}{\small \texttt{-}\textbf{91.9\%}} & \textcolor{darkgreen}{\small \texttt{+}\textbf{1,505\%}} & \textcolor{darkgreen}{\small \texttt{-}\textbf{98.8\%}} & \textcolor{darkgreen}{\small \texttt{-}\textbf{98.3\%}} & \textcolor{darkgreen}{\small \texttt{+}\textbf{1,062\%}} \\
  \midrule
  \multirow{3}{*}{TS-CAN\cite{mttscan}}   & \ding{55} & 7.154 & 9.918 & 0.461 & 9.340 & 12.81 & 0.287 & 12.87 & 20.42 & 0.454 & 8.764 & 12.95 & 0.796 \\
            &\cellcolor{light-gray} \checkmark &\cellcolor{light-gray} 1.401 &\cellcolor{light-gray} 2.740 &\cellcolor{light-gray} 0.965 &\cellcolor{light-gray} 4.253 &\cellcolor{light-gray} 7.496 &\cellcolor{light-gray} 0.783 &\cellcolor{light-gray} 1.078 &\cellcolor{light-gray} 3.204 &\cellcolor{light-gray} 0.990 &\cellcolor{light-gray} \bf{0.519} &\cellcolor{light-gray} \bf{0.741} &\cellcolor{light-gray} \bf{0.999} \\
            & \textsc{Delta} & \textcolor{darkgreen}{\small \texttt{-}\textbf{80.4\%}} & \textcolor{darkgreen}{\small \texttt{-}\textbf{72.4\%}} & \textcolor{darkgreen}{\small \texttt{+}\textbf{109.3\%}} & \textcolor{darkgreen}{\small \texttt{-}\textbf{54.5\%}} & \textcolor{darkgreen}{\small \texttt{-}\textbf{41.5\%}} & \textcolor{darkgreen}{\small \texttt{+}\textbf{172.7\%}} & \textcolor{darkgreen}{\small \texttt{-}\textbf{91.6\%}} & \textcolor{darkgreen}{\small \texttt{-}\textbf{84.3\%}} & \textcolor{darkgreen}{\small \texttt{+}\textbf{118.3\%}} & \textcolor{darkgreen}{\small \texttt{-}\textbf{94.1\%}} & \textcolor{darkgreen}{\small \texttt{-}\textbf{94.3\%}} & \textcolor{darkgreen}{\small \texttt{+}\textbf{25.5\%}} \\
  \midrule
  \multirow{3}{*}{PhysNet\cite{physnet}}  & \ding{55} & 8.738 & 11.23 & 0.186 & 11.88 & 14.98 & 0.062 & 18.46 & 22.53 & 0.407 & 15.09 & 20.18 & 0.490 \\
            &\cellcolor{light-gray} \checkmark &\cellcolor{light-gray} 1.320 &\cellcolor{light-gray} 2.700 &\cellcolor{light-gray} 0.969 &\cellcolor{light-gray} \bf{2.042} &\cellcolor{light-gray} \bf{3.563} &\cellcolor{light-gray} \bf{0.955} &\cellcolor{light-gray} 3.930 &\cellcolor{light-gray} 8.522 &\cellcolor{light-gray} 0.940 &\cellcolor{light-gray} 0.879 &\cellcolor{light-gray} 1.592 &\cellcolor{light-gray} 0.996 \\
            & \textsc{Delta} & \textcolor{darkgreen}{\small \texttt{-}\textbf{84.9\%}} & \textcolor{darkgreen}{\small \texttt{-}\textbf{75.9\%}} & \textcolor{darkgreen}{\small \texttt{+}\textbf{420.4\%}} & \textcolor{darkgreen}{\small \texttt{-}\textbf{82.8\%}} & \textcolor{darkgreen}{\small \texttt{-}\textbf{76.2\%}} & \textcolor{darkgreen}{\small \texttt{+}\textbf{1,440\%}} & \textcolor{darkgreen}{\small \texttt{-}\textbf{78.7\%}} & \textcolor{darkgreen}{\small \texttt{-}\textbf{62.2\%}} & \textcolor{darkgreen}{\small \texttt{+}\textbf{131.0\%}} & \textcolor{darkgreen}{\small \texttt{-}\textbf{94.2\%}} & \textcolor{darkgreen}{\small \texttt{-}\textbf{92.1\%}} & \textcolor{darkgreen}{\small \texttt{+}\textbf{103.1\%}} \\
  \midrule
  \multirow{3}{*}{EFFPhys-C\cite{efficientphys}} & \ding{55} & 5.386 & 7.665 & 0.658 & 9.028 & 11.70 & 0.431 & 11.65 & 17.59 & 0.671 & 6.788 & 10.58 & 0.848 \\
            &\cellcolor{light-gray} \checkmark &\cellcolor{light-gray} 2.379 &\cellcolor{light-gray} 4.450 &\cellcolor{light-gray} 0.904 &\cellcolor{light-gray} 5.716 &\cellcolor{light-gray} 9.684 &\cellcolor{light-gray} 0.624 &\cellcolor{light-gray} 3.864 &\cellcolor{light-gray} 9.010 &\cellcolor{light-gray} 0.926 &\cellcolor{light-gray} 0.900 &\cellcolor{light-gray} 1.738 &\cellcolor{light-gray} 0.995 \\
            & \textsc{Delta} & \textcolor{darkgreen}{\small \texttt{-}\textbf{55.8\%}} & \textcolor{darkgreen}{\small \texttt{-}\textbf{42.0\%}} & \textcolor{darkgreen}{\small \texttt{+}\textbf{37.4\%}} & \textcolor{darkgreen}{\small \texttt{-}\textbf{36.7\%}} & \textcolor{darkgreen}{\small \texttt{-}\textbf{17.2\%}} & \textcolor{darkgreen}{\small \texttt{+}\textbf{44.8\%}} & \textcolor{darkgreen}{\small \texttt{-}\textbf{66.8\%}} & \textcolor{darkgreen}{\small \texttt{-}\textbf{48.8\%}} & \textcolor{darkgreen}{\small \texttt{+}\textbf{37.9\%}} & \textcolor{darkgreen}{\small \texttt{-}\textbf{86.7\%}} & \textcolor{darkgreen}{\small \texttt{-}\textbf{83.6\%}} & \textcolor{darkgreen}{\small \texttt{+}\textbf{17.3\%}} \\
  \midrule
  \multirow{3}{*}{PhysFormer\cite{physformer}} & \ding{55} & 3.536 & 6.746 & 0.904 & 8.445 & 10.62 & 0.438 & 16.94 & 21.62 & 0.642 & 2.577 & 6.587 & 0.926 \\
            &\cellcolor{light-gray} \checkmark &\cellcolor{light-gray} 2.325 &\cellcolor{light-gray} 5.239 &\cellcolor{light-gray} 0.942 &\cellcolor{light-gray} 3.873 &\cellcolor{light-gray} 6.995 &\cellcolor{light-gray} 0.815 &\cellcolor{light-gray} 3.976 &\cellcolor{light-gray} 9.247 &\cellcolor{light-gray} 0.929 &\cellcolor{light-gray} 0.545 &\cellcolor{light-gray} 0.872 &\cellcolor{light-gray} \bf{0.999} \\
            & \textsc{Delta} & \textcolor{darkgreen}{\small \texttt{-}\textbf{34.3\%}} & \textcolor{darkgreen}{\small \texttt{-}\textbf{22.3\%}} & \textcolor{darkgreen}{\small \texttt{+}\textbf{4.2\%}} & \textcolor{darkgreen}{\small \texttt{-}\textbf{54.1\%}} & \textcolor{darkgreen}{\small \texttt{-}\textbf{34.1\%}} & \textcolor{darkgreen}{\small \texttt{+}\textbf{86.1\%}} & \textcolor{darkgreen}{\small \texttt{-}\textbf{76.5\%}} & \textcolor{darkgreen}{\small \texttt{-}\textbf{57.2\%}} & \textcolor{darkgreen}{\small \texttt{+}\textbf{44.7\%}} & \textcolor{darkgreen}{\small \texttt{-}\textbf{78.9\%}} & \textcolor{darkgreen}{\small \texttt{-}\textbf{86.8\%}} & \textcolor{darkgreen}{\small \texttt{+}\textbf{7.9\%}} \\
  \bottomrule
\end{tabular}
\begin{tablenotes}[flushleft]
     \item \textbf{MAE}: Mean Absolute Error, \textbf{RMSE}: Root Mean Square Error, \textbf{$\rho$}: Pearson Correlation Coefficient.
\end{tablenotes}
\end{threeparttable}
}
\end{table*}

Section \ref{sec: dataset}
introduces the   five public datasets used on experiments: RLAP\cite{wang2023physbench}, UBFC-rPPG\cite{ubfcrppg}, PURE\cite{pure}, COHFACE\cite{cohface} and MMPD\cite{tang2023mmpd}. Section \ref{sec: Implementation} details the settings of experiments. Section \ref{sec: cross-dataset} experiments included four commonly used baseline models: TS-CAN\cite{mttscan}, EfficientPhys\cite{efficientphys}, PhysNet\cite{physnet}, and PhysFormer\cite{physformer}, as well as a proposed model called TN-Net. Each model is divided into versions with and without the insertion of the TN module, in order to compare the performance changes brought about by the inclusion of the TN module.

\subsection{Datasets}
\label{sec: dataset}
We selected five representative datasets for our cross-dataset experiments. Below are the detailed descriptions of these datasets:

\textbf{RLAP-rPPG}\cite{wang2023physbench}: Including 58 subjects, each recorded with 4 videos of MJPG compressed video with a resolution of 640x480 for 2 minutes using a low-cost Logitech C930c webcam, the bitrate is approximately 9000 kbps. Physiological signals were recorded by hand using Contec CMS50E blood pulse meter. 

\textbf{UBFC-rPPG}\cite{ubfcrppg}: Including 43 uncompressed RGB videos with a resolution of 640x480, each video is approximately one minute long, recorded at 30fps using a low-cost Logitech C920 HD Pro webcam with a bitrate of 221184 kbps. Physiological signals were recorded by hand using Contec CMS50E blood pulse meter. 

\textbf{PURE}\cite{pure}: Including 59 videos and pulse data of 10 subjects. Videos were recorded using the eco274CVGE robotic camera at a resolution of 640x480 pixels, 30fps in lossless RGB format, the bitrate is 221184 kbps. Physiological signals were recorded by hand using Contec CMS50E blood pulse meter. 

\textbf{COHFACE}\cite{cohface}: Including 160 one-minute videos collected from 40 subjects, with synchronized recordings from three input channels using the BioGraph Infiniti software suite. The videos are provided in MPEG-4 compressed format with a resolution of 640x480 and a frame rate of 20fps, recorded using a Logitech C525 camera at a bitrate of approximately 250 kbps. Physiological signals were recorded using a pulse oximeter from Thought Technologies.

\textbf{MMPD}\cite{tang2023mmpd}: Including 660 one-minute videos from 33 subjects. Videos were recorderd at 30 fps using a Galaxy S22 Ultra smartphone camera at a resolution of 1280x720 pixels, 30fps in H264 compressed format, the bitrate is approximately 10000 kbps. Physiological signals were recorded by hand using HKG-07C+ blood pulse meter. 



\subsection{Experimental Setup}

\label{sec: Implementation}
We replicated four baseline models based on the original paper: TS-CAN\cite{mttscan}, EfficientPhys\cite{efficientphys}, PhysNet\cite{physnet}, and PhysFormer\cite{physformer}, In addition, to validate the effectiveness of the TN module, we propose a novel model named TN-Net. The model structure is similar to TS-CAN\cite{mttscan}, utilizing Temporal s
Shift Module (TSM)\cite{tsm} and convolution as its core components, with a TN module added to the front of each basic block. Since the TN module inherently possesses global correlation, long-distance temporal shifts are unnecessary, so the shift length of TSM is set to 1. Given that the TN module can maintain stability in deep attention mechanisms, its depth is twice that of TS-CAN. Additionally, the feedforward layers at the model's end are simplified, resulting in TN-Net having only half the parameters of TS-CAN, the number of parameters is reduced from 533K to 209K. To streamline the model's preprocessing, TN-Net, inspired by Efficientphys\cite{efficientphys}, employs convolutional self-attention in its attention layers. As a result, it only requires raw frames as input, eliminating the need for pre-processing and functioning as an end-to-end model. To validate the optical model proposed in this study, all differential modules and any neural network layers capable of learning differential features were removed from TN-Net, thus rendering it entirely incapable of extracting differential features. 
TN-Net employs the Adam optimizer and Mean Absolute Error (MAE) loss, with all parameters set to their default values. All models are trained for 20 epochs.

Facial detection was conducted using MediaPipe\cite{lugaresi2019mediapipeframeworkbuildingperception}. During model training, temporal resampling was utilized to modify the heart rate of samples, generating an additional 30\% of training data. This augmented data features a heart rate distribution ranging from 30 BPM to 180 BPM. During testing, a Butterworth filter was employed for band-pass filtering within the 0.5\textasciitilde3 Hz (30\textasciitilde180 BPM) range. 
Subsequently, the Welch method was used to extract the peak of heart rate, utilizing a Hanning window with a size of 256 and zero-padding the data to a length of 3.3 K. Each video is divided into several 15-second clips, with the heart rate calculated for each segment. The average heart rate across all segments is then used as the label or prediction value.

The configuration of the experimental platform, our open-source code, pre-trained weights, including the implementation on the rPPG-Toolbox\cite{liu2022rppgtoolbox} can be found in the supplementary materials.

\subsection{Cross-dataset Testing}
\label{sec: cross-dataset}
Considering the sensitivity of the rPPG task to environmental conditions and equipment, intra-dataset test results often fail to adequately demonstrate the generalizability of the model. Therefore, to reflect the model's performance in real-world scenarios, this paper presents only the results of cross-dataset testing. 

Considering the synchronization and diversity of the datasets, this experiment selects two datasets as training sets, utilizing other datasets as test sets. The first training set, RLAP\cite{wang2023physbench}, features low-compression video and highly synchronized physiological signal labels, making it more conducive to model training. The second training set, MMPD\cite{tang2023mmpd}, includes samples with a variety of different skin tones and diverse movement scenarios, suitable for training models capable of adapting to a wide range of real-world conditions. However, due to substantial environmental noise, higher compression rates, and relatively lower synchronization, successfully training models on MMPD proves to be more challenging.

\begin{figure}[t!]
    \centering
    \includegraphics[width=82.5mm]{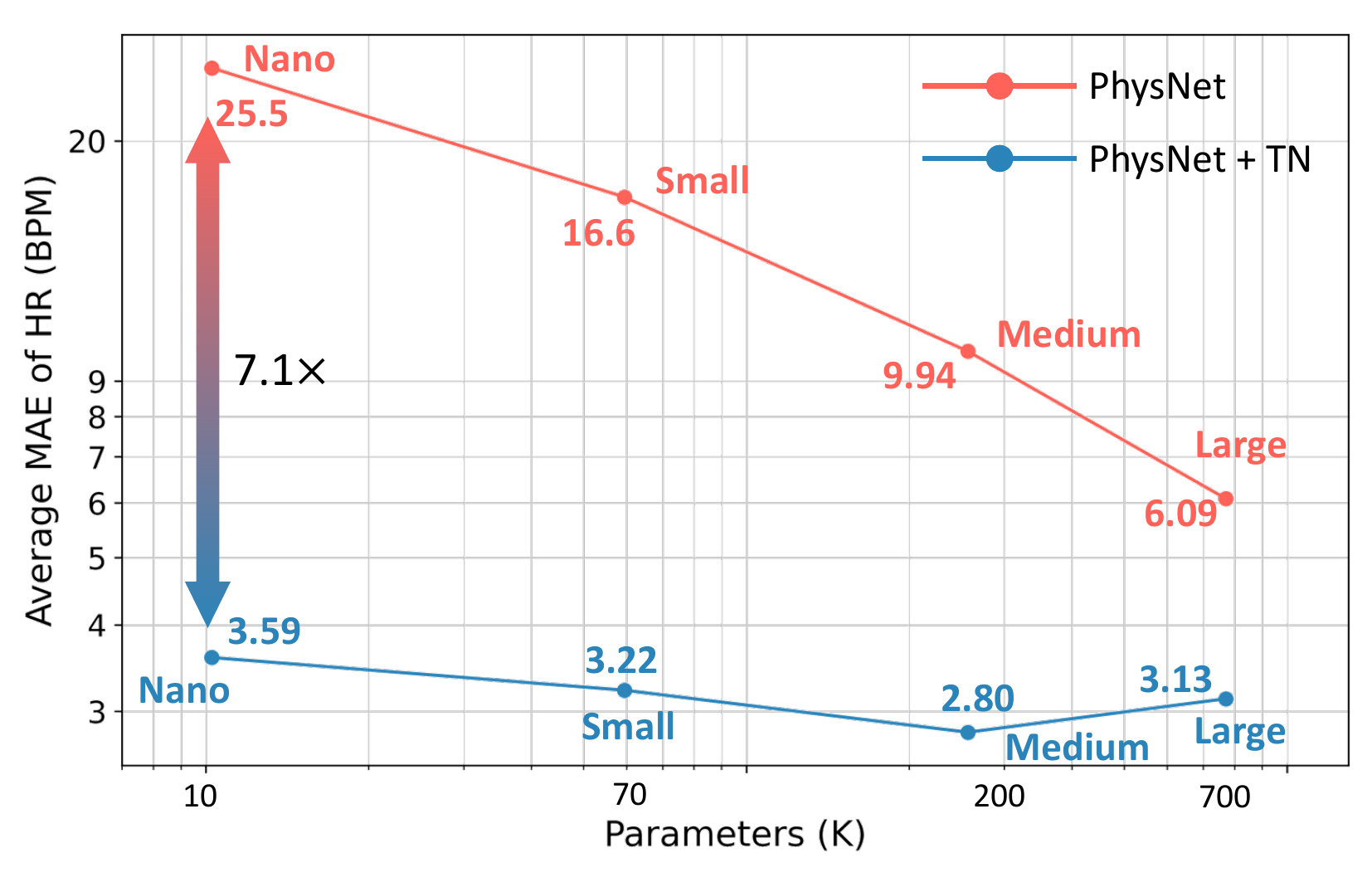}
    \caption{\textbf{The influence of model parameters on the performance of the TN module.} Due to the effective global correlation of TN module, PhysNet can achieve high performance (average MAE 3.59) with only 10K parameters.}
    \label{fig:physnet}
\end{figure}

\subsubsection{Training on RLAP} 

\textbf{TN improves rPPG models' performance trained on RLAP.} The results from training on the RLAP dataset, as shown in Table \ref{resultrlap}, indicate that without the TN module, the four baseline models perform well on the UBFC dataset but exhibit slightly poorer performance on the PURE and COHFACE datasets, and perform poorly on the MMPD dataset. The TN-Net proposed in this study shows almost no effective performance, with outputs that appear nearly random. This is attributable to the removal of any modules capable of capturing differential features, aligning with our optical model, which suggests that existing models rely on differential features. However, upon adding the TN module, the performance of the four baseline models improved significantly, particularly on the PURE and COHFACE datasets, where the average MAE was reduced by 93.0\% and 58.2\%. There are also enhancements in performance on the UBFC\cite{ubfcrppg} dataset, though improvements for MMPD are less substantial. TN-Net fully manifests its potential, achieving optimal results on the MMPD dataset with a MAE of only 6.62, while also performing well on other datasets. The transformation of TN-Net highlights its operation based solely on TN features, and the enhancement in baseline models' performance underscores that TN features surpass differential features in effectiveness.

\subsubsection{Training on MMPD}

\textbf{TN improves rPPG models' performance trained on MMPD.} The results from training on the MMPD\cite{tang2023mmpd} dataset are illustrated in Table \ref{resultmmpd}. The MMPD dataset encompasses a wide array of dynamic scenes and various skin tones, making it more reflective of real-world conditions and thus more valuable as a training dataset. However, due to the high noise levels in MMPD, many neural models find it difficult to converge during training, failing to adequately learn rPPG features, which restricts the practical applicability of MMPD in real-world scenarios. The results demonstrate that baseline models, in the absence of TN features, struggle to train effectively on MMPD, exhibiting low performance in cross-dataset tests, with baseline models' average MAE being 9.79, significantly inferior to results obtained from training on RLAP. However, with the incorporation of TN features, all model performances improved markedly, and the average MAE was reduced to 2.44, indicating that the TN module enhances the convergence and generalization abilities of the models, thereby equipping them with a stronger learning capacity for complex, high-noise datasets.

\section{Discussion} 
In Section \ref{sec: effectiveness}, we discuss the effectiveness of the TN module with cross-dataset results and model-size experiments. Followed by Section \ref{sec: justify}, we explain why TN could work with model attention visualization and prediction analysis. Section \ref{sec: limit} acknowledges the limits of our method and future directions to achieve the full potential of the TN module.

\subsection{Effectiveness of TN Module}
\label{sec: effectiveness}

\textbf{The TN module can enhance model performance in a plug-and-play manner.} As shown in Table \ref{resultrlap}, the TN module can be seamlessly integrated into end-to-end models without additional parameters. It has led to performance improvements ranging from 10.4\% to 98.8\% across various models and datasets. Notably, the best MAE results were recorded on MMPD at 6.62 with TN-Net, on COHFACE at 2.24 with PhysNet-TN, on PURE at 0.232 with TN-Net, and on UBFC at 0.446 with PhyFormer-TN. The only exception occurred when training on RLAP and testing on MMPD using EFFPhys-C, likely due to the oversized but shallow dense layer.

\textbf{The TN module can improve training performance on complex, high-noise real-world datasets.} MMPD \cite{tang2023mmpd}, recognized as a challenging dataset for real-world validation, is difficult to train on due to significant video compression and motion noise. Most previous research either failed to train on the full MMPD dataset \cite{Liu2024SpikingPhysFormerCR} or opted to train on a subset\cite{Zhao_2024_CVPR}. In contrast, our TN module enables all methods to successfully train on the entire MMPD dataset in Table \ref{resultmmpd}, with performance gains ranging from 34.3\% to 98.8\%. Moreover, it helps achieve clinical-level accuracy (MAE below 2) in cross-dataset experiments: 1.114 with TN-Net on RLAP, 2.042 with PhysNet-TN on COHFACE, 0.622 with TN-Net on PURE, and 0.519 with TSCAN-TN on UBFC.

\textbf{TN showed even greater performance gains in smaller models.} Reducing PhysNet’s parameters to only 10K \cite{physnet} resulted in an average MAE increase to 25.5. However, PhysNet-TN maintained a low MAE of 3.59, which is just one-seventh of the standard PhysNet. This performance contrast was further highlighted under varying parameter adjustments, as shown in Figure \ref{fig:physnet}. While the reduction of parameters led to a significant performance drop in standard PhysNet, the decline was minimal in PhysNet-TN. Notably, even with a 98.8\% reduction in parameters (from 770k to 10k), PhysNet-TN still performed better (MAE improved from 6.09 to 3.59). This resilience highlights the TN module’s energy efficiency, making it ideal for lightweight edge devices with minimal impact on performance.

\begin{figure}[t!]
    \centering
    \includegraphics[width=82.5mm]{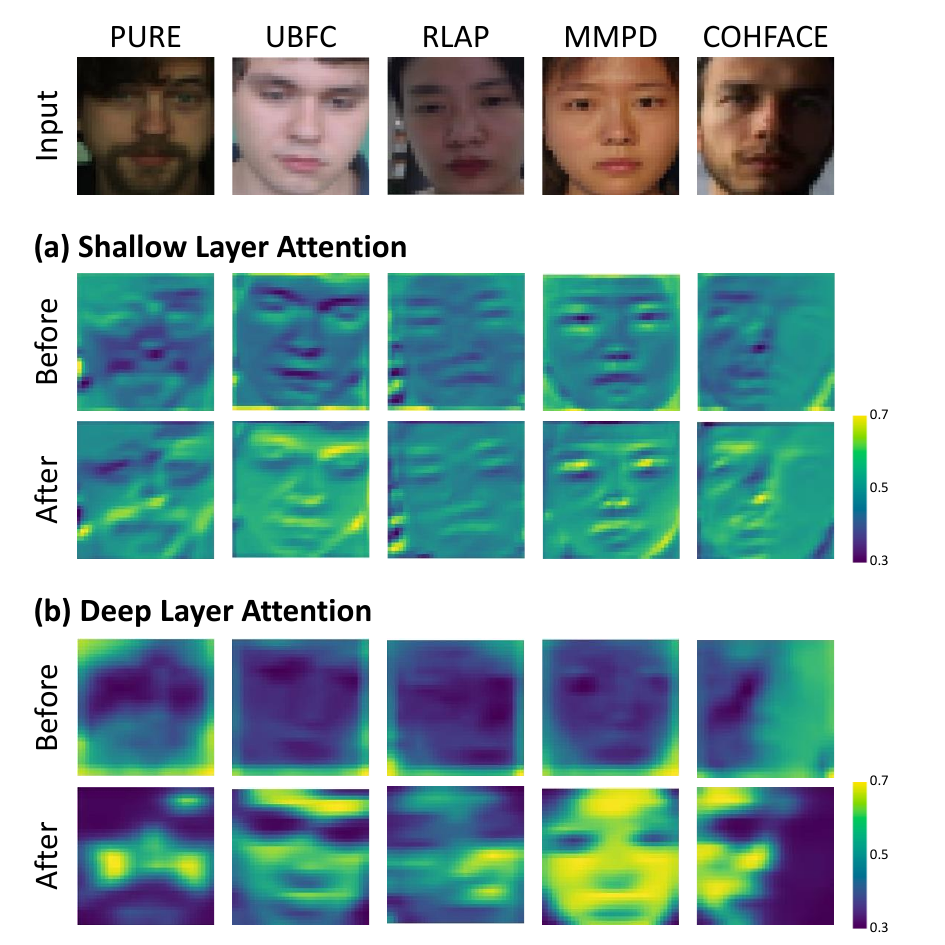}
\caption{\textbf{Visualization of shallow and deep attention in TS-CAN with and without the TN Module.} (a) TS-CAN, with or without the TN module, achieves spatial attention on the facial region. (b) TS-CAN exhibits diminished attention in the absence of the TN module. The comparison shows TN effectively mitigates the issue of vanishing attention in deeper layers.}
    \label{fig:attn}
\end{figure}

\subsection{Justify Why TN Module Work?}
\label{sec: justify}

\textbf{TN establishes long-term correlations across video frames on rPPG models.} As illustrated in Figure \ref{fig:global}, the TN module optimizes this by integrating global information in the temporal dimension. Each frame’s features include a local spatial feature branch and a scale factor matrix branch, which relies on global temporal statistics. While existing models like PhysNet \cite{physnet}, PhysFormer \cite{physformer}, and PhysMamba \cite{physmambayu, Yan2024PhysMambaSS} rely on complex, resource-intensive architectures that require a large number of parameters to manage long-term correlations, our TN module’s dual-branch approach inherently provides these correlations within each frame. This method allows for efficient modeling of extended temporal correlations without additional parameters.

\textbf{TN mitigates the attention vanishment of deep layers of rPPG models.} As depicted in Figure \ref{fig:attn}(b), the deep layers of TS-CAN \cite{mttscan} tend to lose focus on the facial region as the depth increases. In contrast, with the TN module, the same layers maintain a sharp focus on essential facial areas, particularly in skin and brightly lit regions. Previous studies \cite{mttscan, efficientphys} have generally capped model depth at around a dozen layers, as subtle chromatic variations between frames often get lost amid environmental noise, which complicates feature alignment in deeper layers. The TN module, however, facilitates the development of deeper and larger rPPG models by effectively overcoming these challenges.

\textbf{TN amplifies the spatiotemporal features in rPPG models.} Figure \ref{fig:tnvis} presents the visualization of the facial spatiotemporal features amplified by the TN module, with the average RGB color and the model's BVP waveform signal displayed beneath the facial image. The rPPG features in the original image signal are extremely subtle and invisible to the human eye. However, after processing with the TN module, periodic patterns corresponding to the peaks of the BVP waveform can be observed. Due to the enhanced intensity of the features input into the neural network, the features become more robust and less likely to vanish in the deeper layers of the network.


\subsection{Limitations and Future Work}
\label{sec: limit}
We validated the effectiveness of the TN module through experiments and illustrated its interpretability with formulas and visualizations. However, we recognize that this module has limitations in high-noise scenarios. The TN module’s performance relies on the assumption that environmental noise levels are significantly lower than light intensity, as outlined in Inequality \eqref{order}. The MMPD\cite{tang2023mmpd} dataset, which includes data with intense head movements and illumination changes mimicking real-life situations, contains high levels of noise that break this assumption. As indicated by the modest performance improvements in baseline models tested on MMPD (see Table \ref{resultrlap}), the TN module struggles to extract rPPG signals under intense environmental noise.

Future work could focus on effectively managing scenarios with extremely high noise levels in videos. Leveraging the TN module’s ability to significantly reduce deep feature vanishment, we can explore deepening the model layers beyond the current limitation of roughly a dozen layers. TN-Net, although still in its preliminary stages, showcases the potential for expanding both the depth and scale of the model. This approach could enhance the application of rPPG technology in noisy environments, representing a promising research direction.

\section{Conclusions}

By revisiting the Shafer Reflectance Model, this paper proposed a plug-and-play TN module that replaces differential features with temporally normalized features. Experimental results indicate that the TN module could achieve up to a 98.8\% performance improvement on cross-dataset validation. It marks the first success of training on the high-noise MMPD dataset to achieve clinical-level test performance on four other datasets. Moreover, it provides long-term correlations without additional parameters and mitigates the limits of attention vanishing in deeper layers of the model, leading to even greater performance gains in small models.
{
    \small
    \bibliographystyle{plain}
    \bibliography{main}
}

\end{document}